\documentclass[aps,pra,twocolumn,showpacs,floatfix]{revtex4-1}

\usepackage{graphicx}
\usepackage{amsmath}
\usepackage{color}
\usepackage{bm}

\begin{document}

\title{Inelastic losses in radiofrequency-dressed traps for ultracold atoms}

\author{Daniel J. Owens}
\author{Jeremy M. Hutson}
\email{Author to whom correspondence should be addressed:
j.m.hutson@durham.ac.uk} \affiliation{Joint Quantum Centre (JQC)
Durham-Newcastle, Department of Chemistry, Durham University, South Road,
Durham DH1 3LE, United Kingdom.}
\date{\today}

\begin{abstract}
We calculate the rates of inelastic collisions for ultracold alkali-metal atoms
in radiofrequency-dressed traps, using coupled-channel scattering calculations
on accurate potential energy surfaces. We identify an rf-induced loss mechanism
that does not exist in the absence of rf radiation. This mechanism is not
suppressed by a centrifugal barrier in the outgoing channel, and can be much
faster than spin relaxation, which is centrifugally suppressed. We explore the
dependence of the rf-induced loss rate on singlet and triplet scattering
lengths, hyperfine splittings and the strength of the rf field. We interpret
the results in terms of an adiabatic model of the collision dynamics, and
calculate the corresponding nonadiabatic couplings. The loss rate can vary by
10 orders of magnitude as a function of singlet and triplet scattering lengths.
$^{87}$Rb is a special case, where several factors combine to reduce rf-induced
losses; as a result, they are slow compared to spin-relaxation losses. For most
other alkali-metal pairs, rf-induced losses are expected to be much faster and
may dominate.
\end{abstract}

\maketitle

\section{Introduction}

Radiofrequency-dressed traps \cite{Zobay:2001, Zobay:2004} have been widely
used \cite{Garraway:2016} to confine ultracold atoms in complex geometries,
including shells \cite{Colombe:2004} and rings \cite{Heathcote:2008}. These
geometries are valuable in many fields, including condensate splitting and atom
interferometry \cite{Schumm:2005} and the study of low-dimensional quantum
systems \cite{Goebel:thesis:2008}. The atoms are trapped using a combination of
magnetic and radiofrequency (rf) fields, and are confined in an adiabatic
potential obtained by diagonalizing a simple Hamiltonian in a basis set of
rf-dressed atomic states. Radiofrequency dressing has also been used to form
new structures in optical lattices \cite{Lundblad:2008, Shotter:2008}.

There are various sources of losses of atoms from rf-dressed traps. The
rf-dressed state that is adiabatically trapped is not the lowest that exists,
and Burrows {\em et al.}\ \cite{Burrows:2017} have considered one-body losses
due to nonadiabatic transitions to lower states. Such losses may be made
acceptably small by avoiding very low rf coupling strengths. However, the
presence of rf radiation introduces additional loss mechanisms that are not
present for atoms in a purely magnetic trap, due to rf-induced inelastic
collisions.

Tscherbul {\em et al.}\ \cite{TVTscherbul:rf:2010} have developed a
coupled-channel theory of atomic collisions in rf fields and applied it to
rf-induced resonances in $^{87}$Rb. Hanna {\em et al.}\ \cite{Hanna:2010}
developed an approach based on multichannel quantum defect theory, and also
explored rf-induced resonances in $^{87}$Rb and $^6$Li. However, Hanna {\em et
al.}\ stated that calculations with their method were impractical close to
atomic rf transitions, which is precisely the case that is required to
investigate collisions of atoms in rf-dressed traps. Owens {\em et al.}\
\cite{Owens:rf:2016} have shown that rf dressing can be used to create new
Feshbach resonances at desired magnetic fields that are convenient for atomic
cooling. These resonances may be valuable for molecule formation, particularly
in heteronuclear systems.

Most experimental work with rf-dressed traps to date has used $^{87}$Rb atoms.
However, there is considerable interest in extending this to other atomic
species. In this paper we use coupled-channel calculations to explore the rates
of rf-induced inelastic collisions theoretically. We show that $^{87}$Rb is a
special case, where the losses induced by rf radiation are very small. Most
other alkali-metal atoms may be expected to have much faster rf-induced
collisional losses.

\section{Methods}

We use the convention that lower-case quantum numbers refer to individual atoms
and upper-case quantum numbers refer to the colliding pair. In the absence of
fields, each atom is described by its electron spin $s=1/2$ and nuclear spin
$i$, which couple to form a resultant $f$. In a magnetic field $B$, each state
splits into components labeled by $m_f=m_s+m_i$, where each $m$ is the
projection of the corresponding quantity on the magnetic field axis $Z$. The
Hamiltonian for each atom is
\begin{equation}
\hat h = \zeta
\hat i \cdot \hat s + \left( g_S \hat s_z +
g_i \hat i_{z}\right) \mu_{\rm B} B,
\label{eq:mon}
\end{equation}
where $\zeta$ is the hyperfine coupling constant and $g_S$ and $g_i$ are
electron and nuclear spin $g$-factors with the sign convention of Arimondo {\em
et al.}\ \cite{Arimondo:1977}. At the low magnetic fields considered here, $f$
is nearly conserved but $m_s$ and $m_i$ are not.

To incorporate the effects of rf radiation on a single atom, we use a basis set
of photon-dressed functions in an uncoupled representation, $|s m_s\rangle |i
m_i\rangle |N M_N\rangle$, where $N$ is the photon number with respect to the
average photon number $N_0$. In the present work we focus on circularly
polarized radiation, with either $M_N=N$ (right-circularly polarized,
$\sigma_+$) or $M_N=-N$ (left-circularly polarized, $\sigma_-$). In the present
work we consider $\sigma_-$polarization, with $\bm{B}(t)=B_{\rm rf} [\hat e_x
\cos 2\pi\nu t - \hat e_y \sin 2\pi\nu t]$, where $\hat e_x$ and $\hat e_y$ are
unit vectors along the $X$ and $Y$ axes. The Hamiltonian of the rf field is
\begin{equation}
\hat h_{\rm rf} = h\nu(\hat{a}_-^\dagger\hat{a}_--N_0),
\end{equation}
where $\hat{a}_-$ and $\hat{a}_-^\dagger$ are photon annihilation and creation
operators for $\sigma_-$ photons. Its interaction with an atom is
\begin{equation}
\hat h_{\rm rf}^{\rm int} =\frac{\mu_{\rm B} B_{\rm rf}}{2\sqrt{N_0}}
\left[(g_S\hat s_+ + g_i\hat i_+)\hat a_-^\dagger + (g_S\hat s_- + g_i\hat
i_-)\hat a_-\right], \label{eq:rf}
\end{equation}
where $B_{\rm rf}$ is the amplitude of the oscillating magnetic field, $\hat
s_+$ and $\hat s_-$ are raising and lowering operators for the electron spin
and $\hat i_+$ and $\hat i_-$ are the corresponding operators for the nuclear
spin. For $\sigma_+$ polarization, $\hat a_+$ replaces $\hat a_-^\dagger$ and
$\hat a_+^\dagger$ replaces $\hat a_-$ in Eq.\ \eqref{eq:rf}. For $\sigma_X$
radiation with rf field $\bm{B}(t)=B_{\rm rf} \cos 2\pi\nu t$, both $\sigma_+$
and $\sigma_-$ coupling terms are present, renormalized by 1/2
\footnote{Different papers use Hamiltonians corresponding to different
definitions of $B_{\rm rf}$. In particular, our previous work
\cite{Owens:rf:2016} used the same definition of $B_{\rm rf}$ as here for
$\sigma_X$ but a different normalization for $\sigma_+$ and $\sigma_-$. To
match the present definition, the values of $B_{\rm rf}$ given in ref.\
\cite{Owens:rf:2016} must be divided by $\sqrt{2}$ for $\sigma_+$ and
$\sigma_-$ polarization. Note that Hanna \emph{et al.}\ \cite{Hanna:2010} use
spin raising and lowering operators normalized by $1/\sqrt{2}$ with respect to
ours.}.

\begin{figure}[t]
\includegraphics[width=\columnwidth]{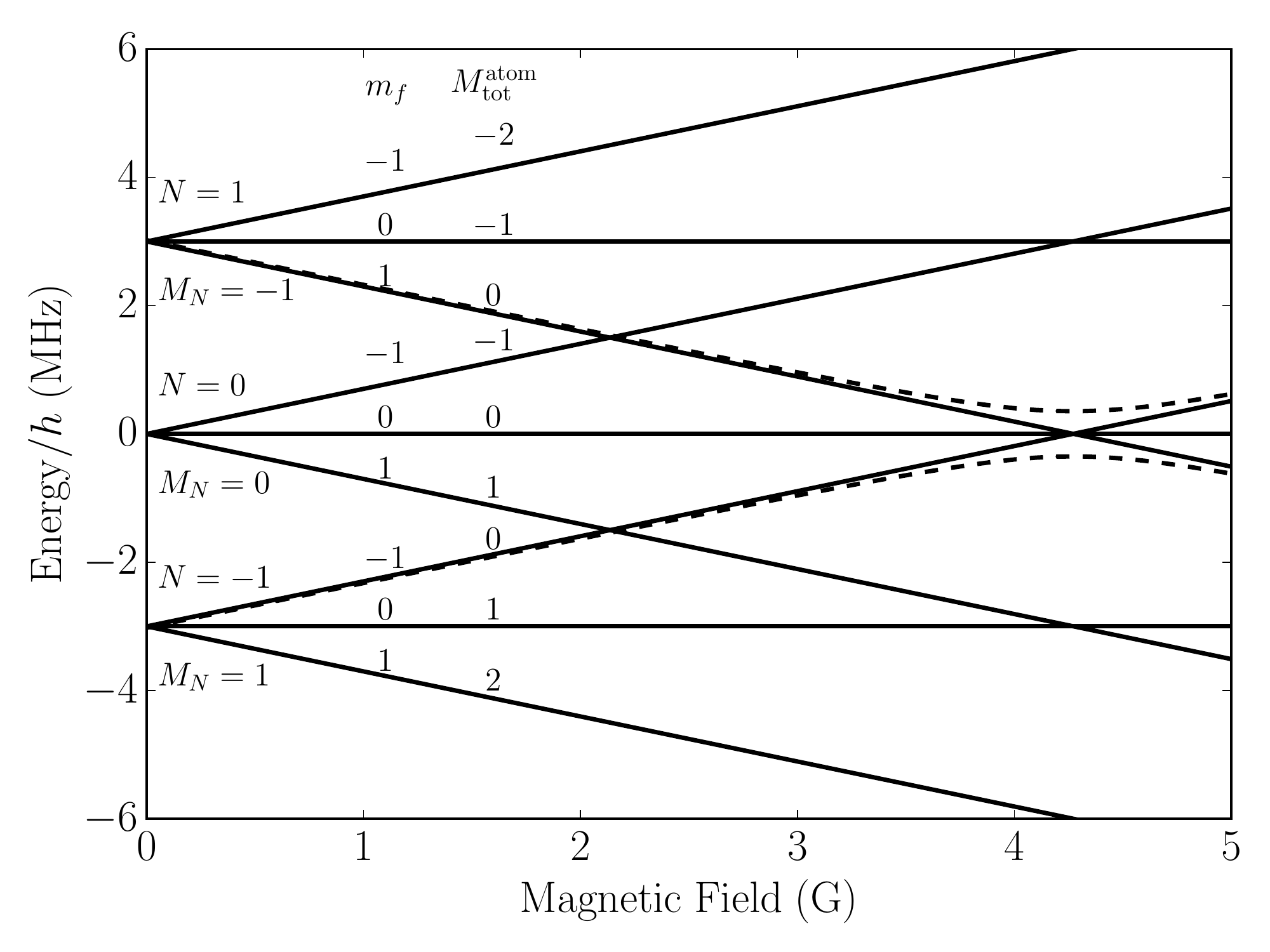}
\caption{rf-dressed atomic levels of $f=1$ states of $^{87}$Rb for frequency
$3.0$ MHz and photon numbers $N=-1$, 0 and 1, shown with respect to the energy
of the $f=1$, $m_f=0$ state for $N=0$. Solid lines show levels for zero rf
intensity and dashed curves show levels for $B_{\rm rf}=0.5$\,G with $M_{\rm
tot}^{\rm atom}=0$. Atoms can be trapped at the minimum in the upper dashed
curve.} \label{fig:Rbatomic}
\end{figure}

If couplings involving the photon annihilation and creation operators are
neglected, states with different $m_f$ values and different photon numbers $N$
cross as a function of magnetic field. For example, for $^{87}$Rb, with
$i=3/2$, and 3.0 MHz radiation, the $(f,m_f,N)=(1,+1,1)$, $(1,0,0)$ and
$(1,-1,-1)$ states all cross near $B=4.27$~G, as shown in Fig.\
\ref{fig:Rbatomic}. If the radiation has $\sigma_-$ polarization, these three
states all have the same total projection quantum number $M_{\rm tot}^{\rm atom}$,
and are coupled by the interaction (\ref{eq:rf}), so the
triple crossing becomes an avoided crossing; for $B_{\rm rf}=0.5$~G, the
minimum separation between the states is $h\times 0.35$~MHz. Ultracold atoms
in the uppermost state can be trapped in the vicinity of the avoided crossing.
These atoms are in a state whose character is principally $(1,+1,1)$ on the
low-field side of the crossing, but is $(1,-1,-1)$ on the high-field side and a
complicated mixture of all three states close to the crossing itself.

We carry out quantum scattering calculations of collisions between pairs of
atoms in rf-dressed states. The Hamiltonian for the colliding pair is
\begin{align}
\frac{\hbar^2}{2\mu} \left[-R^{-1} \frac{d^2}{dR^2} R + \frac{\hat
L^2}{R^2} \right] &+ \hat V(R) + \hat h_1 + \hat h_2 \nonumber\\
&+ \hat h_{\rm rf} + \hat h_{\rm rf,1}^{\rm int} + \hat h_{\rm rf,2}^{\rm int},
\label{eq:ham-pair}
\end{align}
where $\mu$ is the reduced mass, $\hat L^2$ is the operator for the
end-over-end angular momentum of the two atoms about one another, and $\hat
V(R)$ is the interaction operator,
\begin{equation}
{\hat V}(R) = \hat V^{\rm c}(R) + \hat V^{\rm d}(R).
\label{eq:V-hat}
\end{equation}
Here $\hat V^{\rm c}(R)=V_0(R)\hat{\cal{P}}^{(0)} + V_1(R)\hat{\cal{P}}^{(1)}$
is an isotropic potential operator that depends on the electronic potential
energy curves $V_0(R)$ and $V_1(R)$ for the singlet and triplet electronic
states and $\hat V^{\rm d}(R)$ is a relatively weak anisotropic operator that
arises from the combination of spin dipolar coupling at long range and
second-order spin-orbit coupling at short range. The singlet and triplet
projectors $\hat{ \cal{P}}^{(0)}$ and $\hat{ \cal{P}}^{(1)}$ project onto
subspaces with total electron spin quantum numbers 0 and 1 respectively. The
potential curves for the singlet and triplet states are taken from ref.\
\cite{Strauss:2010} for Rb$_2$ and from ref.\ \cite{Falke:2008} for K$_2$.

The basis set of photon-dressed functions for a pair of atoms is
\begin{equation}
|s_1 m_{s1}\rangle |i_1 m_{i1}\rangle |s_2 m_{s2}\rangle |i_2 m_{i2}\rangle |L
M_L\rangle |N M_N\rangle,
\end{equation}
where $L$ is the angular momentum for relative motion and $M_L$ is its
projection onto $Z$. The basis set is symmetrized to take account of exchange
symmetry. The matrix elements of the Hamiltonian in this basis set have been
given in the Appendix of ref.~\cite{Hutson:Cs2-note:2008}, except for the rf
terms, which involve raising and lowering operators with non-zero matrix
elements
\begin{align}
\langle s m_s\pm1 | \hat s_\pm | s m_s \rangle &= [s(s+1)-m_s(m_s\pm1)]^\frac{1}{2}; \\
\langle i m_i\pm1 | \hat i_\pm | i m_i \rangle &= [i(i+1)-m_i(m_i\pm1)]^\frac{1}{2}
\end{align}
and photon creation and annihilation operators with non-zero matrix elements
\begin{align}
\langle N+1 M_N\pm1 | \hat a_\pm^\dagger | N M_N \rangle &= (N_0 + N + 1)^\frac{1}{2};\\
\langle N-1 M_N\mp1 | \hat a_\pm | N M_N \rangle &= (N_0 + N)^\frac{1}{2};\\
\langle N M_N | \hat{a}_\pm^\dagger \hat{a}_\pm | N M_N \rangle &= N_0 + N.
\end{align}
We assume $N_0\gg N$, so that the matrix elements of $\hat a_\pm^\dagger$ and
$\hat a_\pm$ cancel with the factor $N_0^{1/2}$ in the denominator of Eq.\
\eqref{eq:rf}.

The only conserved quantum numbers in a collision are parity $(-1)^L$ and the
total projection $M_{\rm tot}=M_F+M_L+M_N$, where $M_F=m_{f1}+m_{f2}$. Our
basis set includes all possible values of $m_{s1}$, $m_{s2}$, $m_{i1}$,
$m_{i2}$ and $M_L$, for each value of $L$ and $M_N$ that give the required
parity and $M_{\rm tot}$. The values of $L$, $N$ and $M_N$ are limited by $L\le
L_{\rm max}$, $|N|\le N_{\rm max}$ and $|M_N|\le N_{\rm max}$; the values used
for $L_{\rm max}$ and $N_{\rm max}$ will be described for each set of
calculations.

Expanding the scattering wavefunction in the basis set described above produces
a set of coupled equations in the interatomic distance coordinate $R$. The
number of coupled equations varies from 30 to 208. These equations are solved
using the MOLSCAT package \cite{molscat:2017}. In the present work we use the
hybrid log-derivative propagator \cite{Alexander:1987} to propagate the coupled
equations from short range out to $R_{\rm max}=15,000$~bohr. MOLSCAT applies
scattering boundary conditions to extract the scattering $S$ matrix, and then
obtains the complex energy-dependent scattering length
$a(E,B)=\alpha(E,B)-i\beta(E,B)$ from the identity \cite{Hutson:res:2007}
\begin{equation}
a(E,B) = \frac{1}{ik} \left(\frac{1-S_{00}(E,B)}{1+S_{00}(E,B)}\right),
\label{eq:scat-length}
\end{equation}
where $k^2=2\mu E/\hbar^2$ and $S_{00}(E,B)$ is the diagonal S-matrix element
in the incoming s-wave channel. This is constant as $E\rightarrow 0$, where it
reduces to the usual zero-energy scattering length in the absence of inelastic
collisions. For s-wave collisions (incoming $L=0$), the rate coefficient for
inelastic loss is \cite{Cvitas:li3:2007}
\begin{equation}
k_2(E,B) = \frac{2hg_\alpha\beta(E,B)}{\mu\left[1+k^2|a(E,B)|^2+2k\beta(E,B)\right]},
\end{equation}
where $g_\alpha$ is 2 for identical bosons and 1 for distinguishable particles.
This is independent of energy in the limit $E\rightarrow 0$, but resonant peaks
are moderated by the $k^2|a|^2$ term in the denominator at the collision energy
of 1\,$\mu{\rm K}\ \times k_{\rm B}$ used in the present calculations.

\begin{figure}[t]
\includegraphics[width=\columnwidth]{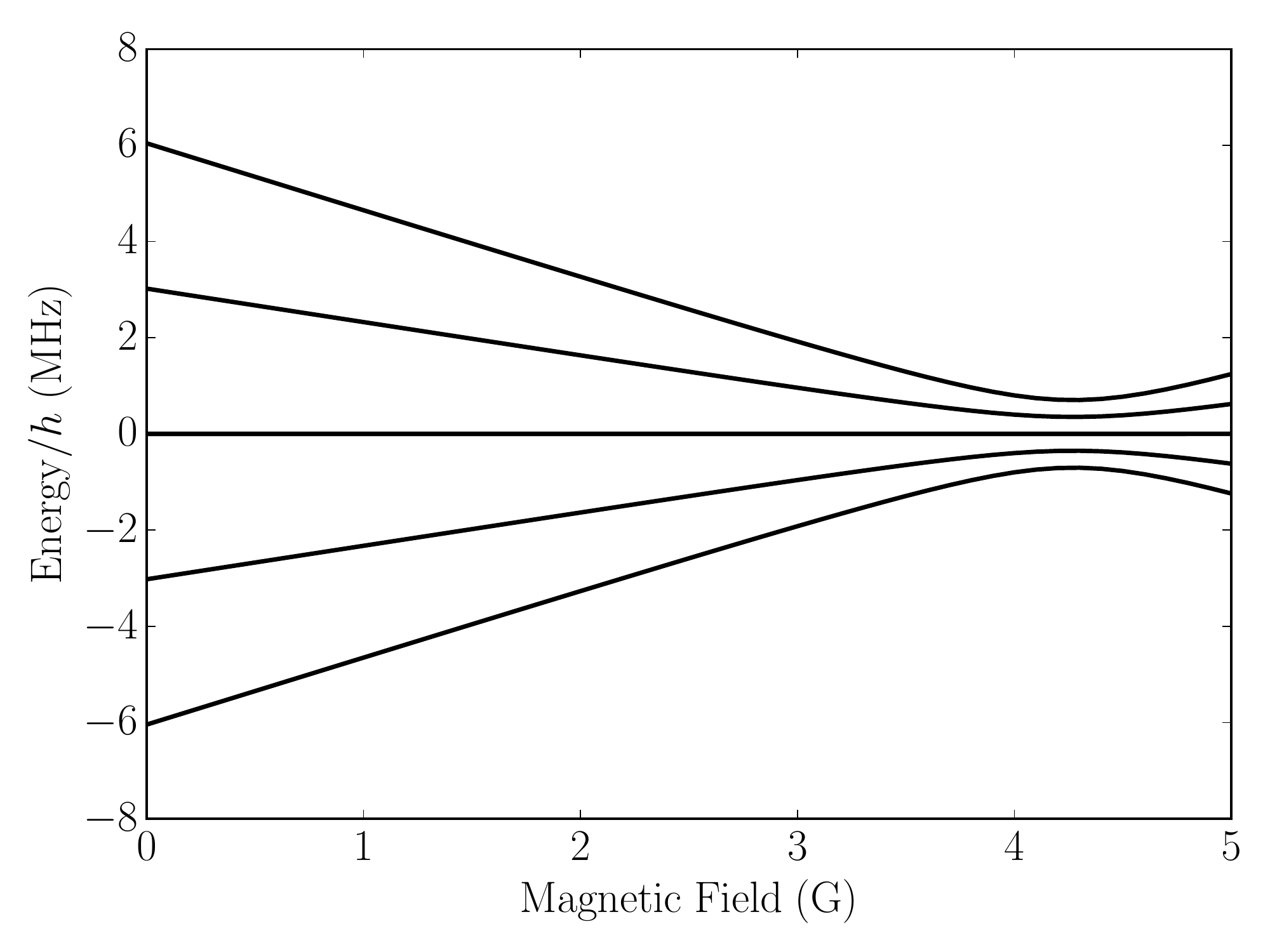}
\caption{The rf-dressed atomic thresholds of $^{87}$Rb+$^{87}$Rb for $f=1$ and
$M_{\rm tot}=0$. The corresponding thresholds of $^{39}$K+$^{39}$K are almost
identical. The rf-induced collisions that cause trap loss are from the
uppermost of these thresholds to all the lower ones. The thresholds are
calculated for $\nu=3.0$ MHz and $B_{\rm rf}=0.5$ G. Near zero magnetic field
the thresholds can be labeled from top to bottom as $M_F=-2,-1,0,+1,+2$ with
$N=-2,-1,0,+1,+2$, respectively.} \label{fig:87Rb2thresh}
\end{figure}

\section{Results}

In this section we present the results of coupled-channel scattering
calculations for rf-dressed states of $^{39}$K and $^{87}$Rb in the vicinity of
an rf-dressed trap. Since $^{87}$Rb is a special case with highly atypical
properties, we consider first the more typical case of $^{39}$K, which (like
$^{87}$Rb) has nuclear spin $i=3/2$ and a hyperfine ground state with $f=1$.

To describe an rf-dressed trap for $f=1$ atoms requires a minimum of 3 rf-free
states with $(f,m_f,N)=(1,+1,1)$, $(1,0,0)$ and $(1,-1,-1)$, as shown in Fig.\
\ref{fig:Rbatomic}. To describe a {\em pair} of such states requires photon
numbers $N$ from $-2$ to 2. For rf field amplitude $B_{\rm rf}=0.5$\,G, this
produces atomic collision thresholds as shown for $^{87}$Rb+$^{87}$Rb in Fig.\
\ref{fig:87Rb2thresh}. The thresholds for $^{39}$K+$^{39}$K are almost
identical. Pairs of atoms are trapped at the highest of the 5 thresholds shown,
and can undergo inelastic collisions to produce atoms at the lower thresholds.
Such inelastic collisions release kinetic energy of at least $h\times 0.25$~MHz
$\approx$ $k_{\rm B}\times 12.5\ \mu$K, and the recoil will usually eject both
collision partners from the trap.

\begin{figure}[t]
\includegraphics[width=\columnwidth]{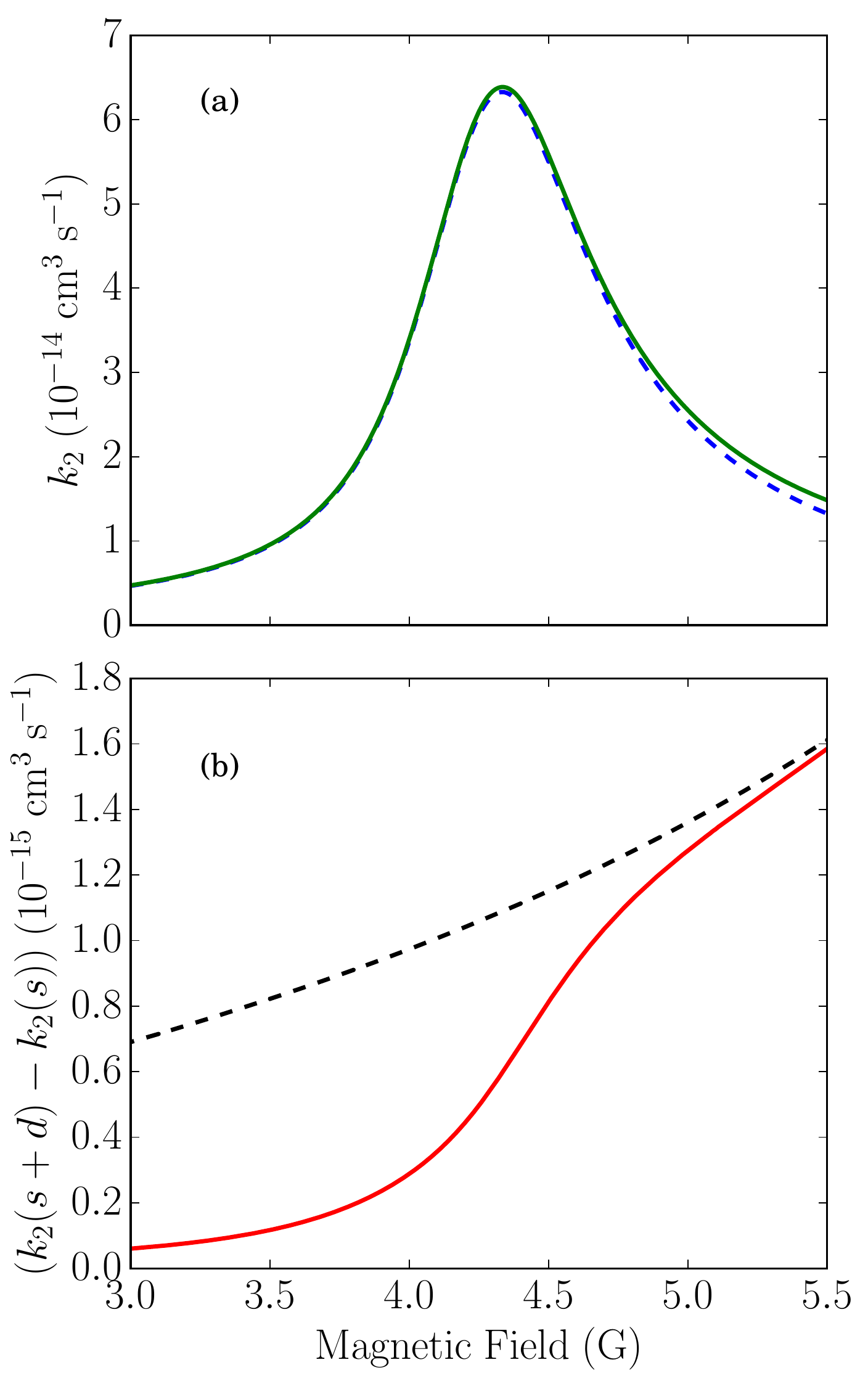}
\caption{(a) Rate coefficient for inelastic loss of adiabatically trapped
$^{39}$K+$^{39}$K as a function of magnetic field with $\nu = 3.0$ MHz and
$B_{\rm rf} = 0.5$ G, from calculations with $L_{\rm max}=2$ (solid, green) and
$L_{\rm max}=0$ (dashed, blue). (b) Contribution from rf-modified
spin-relaxation collisions, obtained from the difference between the $L_{\rm
max}=0$ and $L_{\rm max}=2$ results (red, solid) compared with rf-free
spin-relaxation for $(f,m_f)=(1,-1)$ atoms (black, dashed).}
\label{fig:39K_5e-1G-rf-dressed-loss_lmax0vs2}
\end{figure}

\subsection{Inelastic collisions of rf-dressed $^{39}$K} \label{sec:k2}

The inelastic collision rates for $^{39}$K+$^{39}$K at $B_{\rm rf}=0.5$\,G are
shown in Fig.\ \ref{fig:39K_5e-1G-rf-dressed-loss_lmax0vs2}(a), as a function
of magnetic field across the trap. The solid line shows the inelastic rate from
calculations with $L_{\rm max}=2$, while the dashed line shows the rate from
simplified (and computationally far cheaper) calculations with $L_{\rm max}=0$.
Both calculations use photon numbers $-2\le N\le 2$, and adding additional
values of $N$ makes no further difference to the results.

\begin{figure}[t]
\includegraphics[width=\columnwidth]{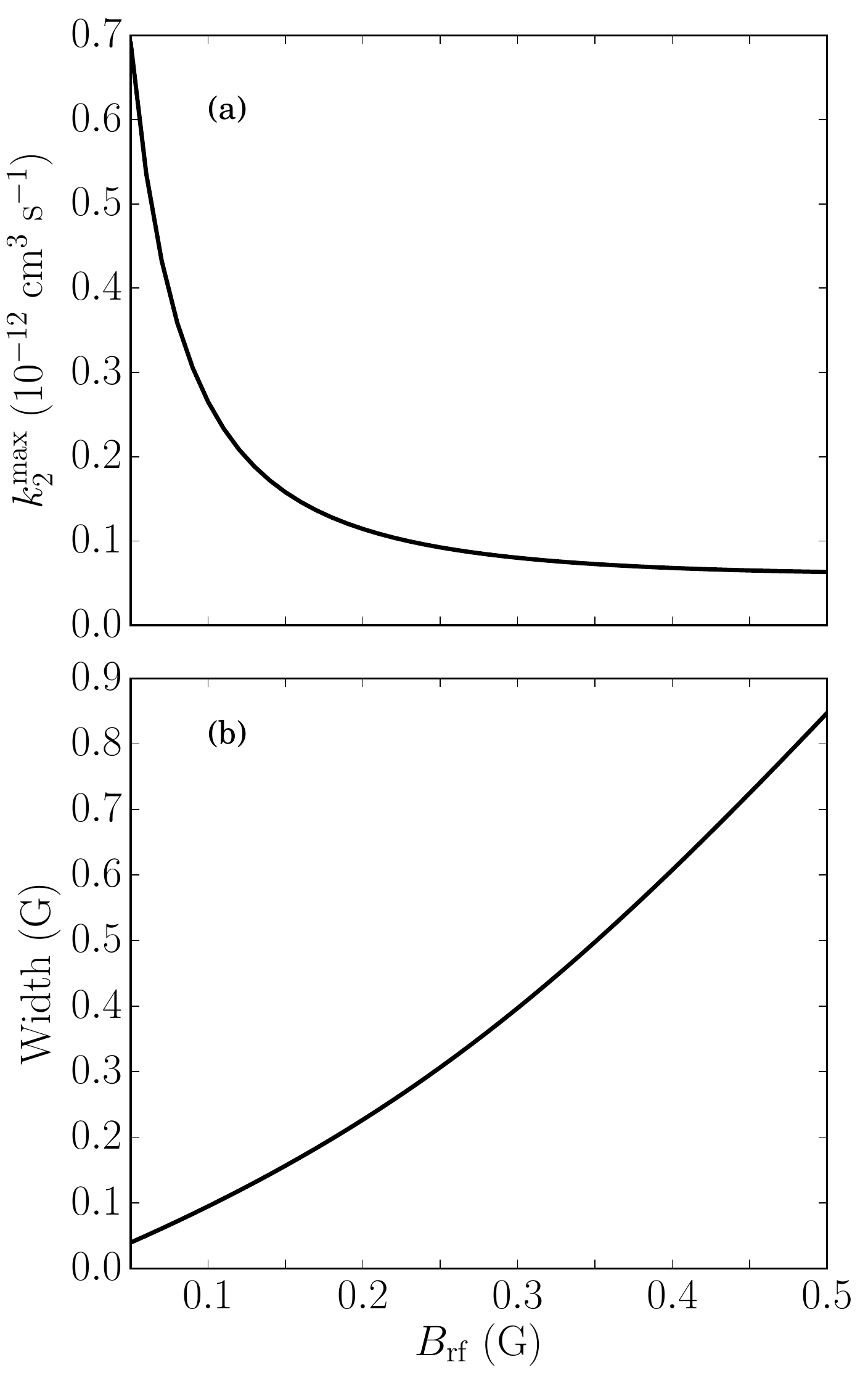}
\caption{Height and FWHM width of the peak in inelastic rate coefficient for
$^{39}$K+$^{39}$K, as a function of rf amplitude $B_{\rm rf}$.}
\label{fig:39K_5e-1G-rf-dressed-loss_peak-height-width}
\end{figure}

The main source of inelasticity in $^{39}$K+$^{39}$K collision exists even for
$L_{\rm max}=0$. It arises from collisions that conserve $m_{f1}+m_{f2}+M_N$
and thus do not change $M_L$; we refer to these are rf-induced collisions.
Since $L$ does not need to change, there is no centrifugal barrier in the
outgoing channel and no centrifugal suppression of the inelastic rate. For
$B_{\rm rf}=0.5$\,G, the loss rate peaks at $k_2^{\rm max}=6.33 \times
10^{-14}$ cm$^3$\,s$^{-1}$ ($\beta=0.015$ bohr) near the trap center and dies
off on either side. However, the peak is a strong function of $B_{\rm rf}$.
Figure \ref{fig:39K_5e-1G-rf-dressed-loss_peak-height-width} shows the height
$k_2^{\rm max}$ and full width at half maximum (FWHM) of the peak as a function
of $B_{\rm rf}$, obtained from calculations with $L_{\rm max}=0$; the peak
width increases as $B_{\rm rf}$ increases, but the peak height decreases. The
width increases faster than linearly with $B_{\rm rf}$; although the range of
$B$ across which the atomic states are strongly mixed by rf dressing is linear
in $B_{\rm rf}$, the kinetic energy released also depends on $B_{\rm rf}$ and
this affects the inelastic cross sections in a complicated way. The peak cross
section decreases as the kinetic energy release increases.

For $^{39}$K+$^{39}$K, the inelastic rates are fairly similar for $L_{\rm
max}=2$ and $L_{\rm max}=0$. The small difference arises because, even in the
absence of rf radiation, atoms in $f=1,m_f<1$ may undergo spin-relaxation
collisions to produce atoms in lower magnetic sublevels. Such collisions are
driven only by the weak anisotropic part of the interaction, $V^{\rm d}(R)$ in
Eq.\ \eqref{eq:V-hat}. Since they change $M_F=m_{f1}+m_{f2}$, and $M_F+M_L$
must be conserved, they must also change $M_L$. For s-wave collisions, $L$ is
initially zero, so changing $M_L$ requires a final state with $L>0$, which must
have $L\ge2$ to conserve parity. The rates of spin-relaxation collisions are
therefore suppressed because the products are trapped inside an $L=2$
centrifugal barrier, which has height $k_{\rm B}\times 1.5$~mK for
$^{39}$K+$^{39}$K. Figure \ref{fig:39K_5e-1G-rf-dressed-loss_lmax0vs2}(b) shows
the {\em difference} between the $L_{\rm max}=2$ and $L_{\rm max}=0$ results in
Fig.\ \ref{fig:39K_5e-1G-rf-dressed-loss_lmax0vs2}(a) and compares it with the
rate of spin-relaxation collisions from an rf-free calculation for two atoms
initially in the $(f,m_f)=(1,-1)$ state. It may be seen that the difference
approaches the rf-free spin-relaxation rate at high magnetic field, where the
adiabatically trapped state is principally $(1,-1)$. However, it decreases to
zero at low magnetic field, where the trapped state is principally $(1,1)$,
which is the rf-free ground state and cannot undergo inelastic collisions. At
the trap center the rf-modified spin-relaxation rate is about half its rf-free
value.

\begin{figure}[t]
\includegraphics[width=\columnwidth]{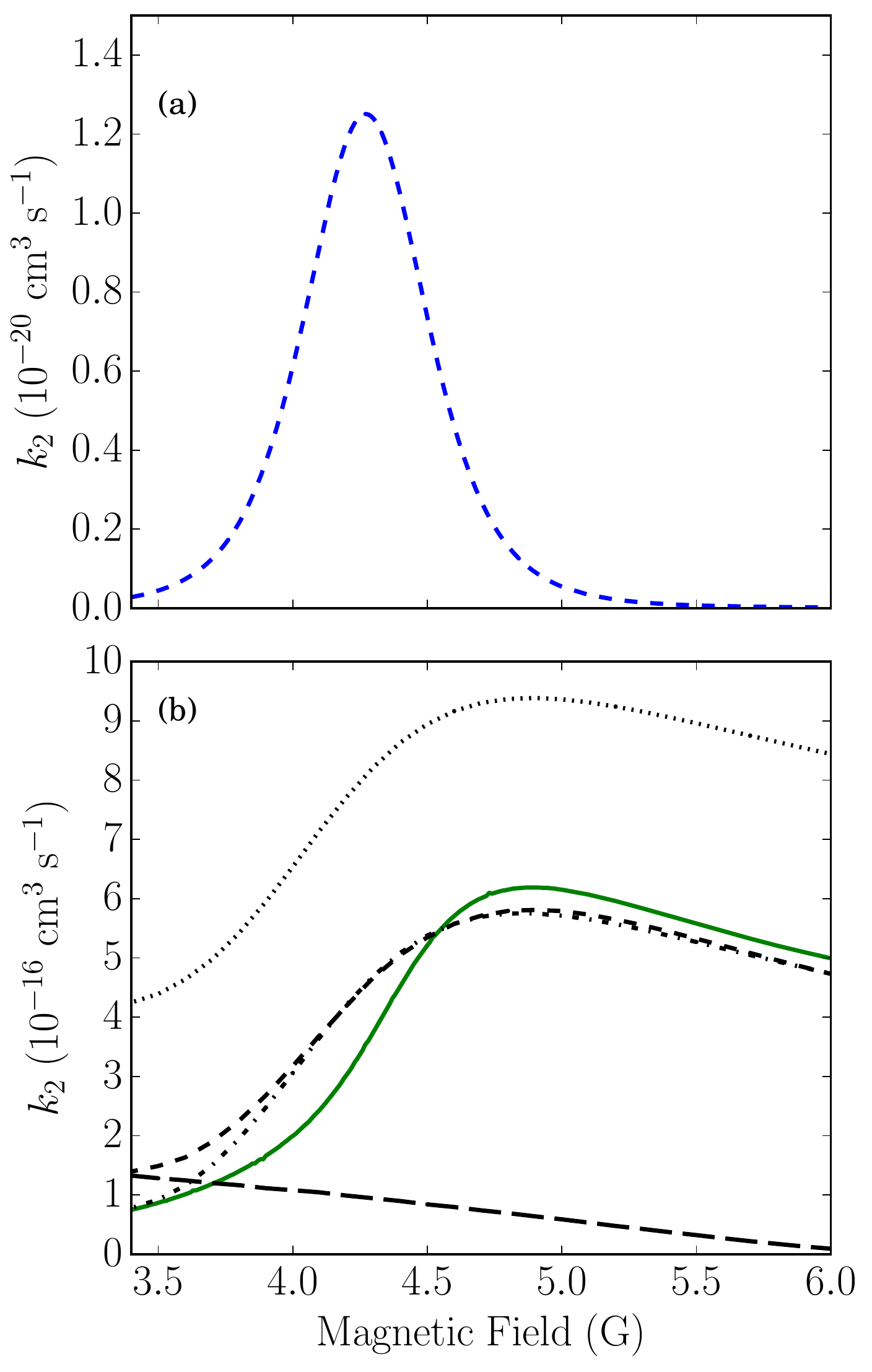}
\caption{Rate coefficient for inelastic loss for adiabatically trapped
$^{87}$Rb in $f=1$ as a function of magnetic field with $\nu = 3.0$ MHz and
$B_{\rm rf} = 0.5$ G. (a) Calculation using $L_{\rm max}=0$. (b) Calculation
including spin relaxation, using $L_{\rm max}=2$ (solid green line). Rate
coefficients for rf-free spin relaxation are shown as dashed lines for
$(1,-1)$+$(1,-1)$, dashed-dotted lines for $(1,-1)$+$(1,0)$, dotted lines
for$(1,0)$+$(1,0)$ and long dashed lines for $(1,0)$+$(1,1)$.}
\label{fig:87Rb2-Brf0.5}
\end{figure}

\begin{figure}[t]
\includegraphics[width=1\columnwidth]{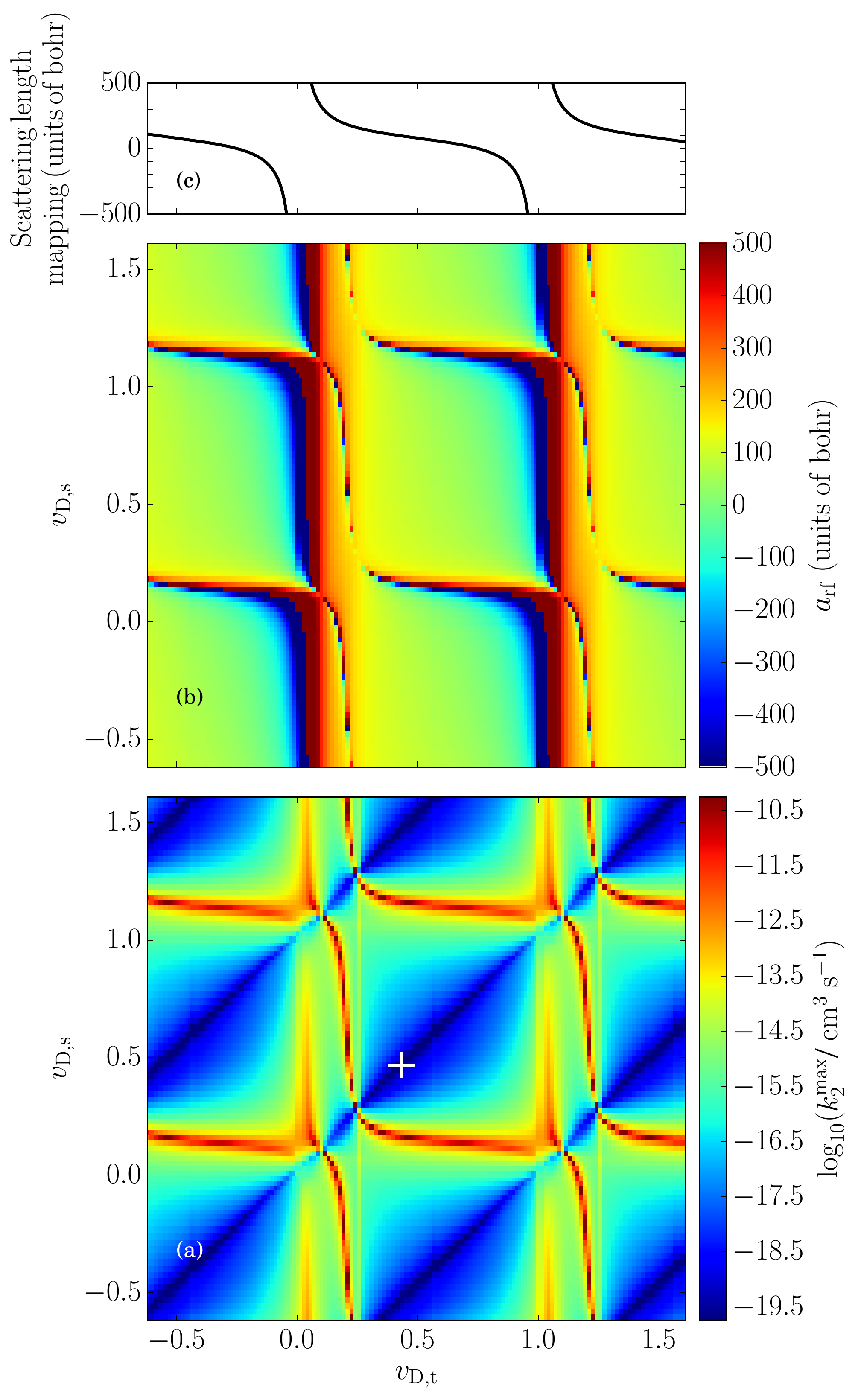}
\caption{Contour plots of the dependence of collision properties on the
fractional part of $v_{\rm D}$ for the singlet and triplet states, for
adiabatically trapped $^{87}$Rb in $f=1$ with $\nu = 3.0$ MHz and $B_{\rm rf} =
0.5$~G. (a) Rate coefficient for rf-induced loss at the trap center, $k_2^{\rm
max}$; (b) Real part of scattering length $a_{\rm rf}$. (c) Mapping between
$v_{\rm D}$ and the singlet and triplet scattering lengths for $^{87}$Rb,
according to Eq.\ \ref{eq:avD}. } \label{fig:contour}
\end{figure}

\subsection{Inelastic collisions of rf-dressed $^{87}$Rb}

Figure \ref{fig:87Rb2-Brf0.5} show the calculated inelastic rate constant as a
function of magnetic field for $^{87}$Rb, for the same rf frequency and field
strength as Fig.\ \ref{fig:39K_5e-1G-rf-dressed-loss_lmax0vs2}. Figures
\ref{fig:87Rb2-Brf0.5}(a) and \ref{fig:87Rb2-Brf0.5}(b) show calculations with
$L_{\rm max}=0$ and 2, respectively. In this case the rate coefficient for
rf-induced loss (with $L_{\rm max}=0$) reaches a maximum of only
$k_2^{\rm max}=1.25 \times 10^{-20}$ cm$^3$\,s$^{-1}$ ($\beta=6.47
\times 10^{-8}$~bohr) at $B=4.2713$~G (the trap center). This is more than 6
orders of magnitude slower than for $^{39}$K$_2$, and 4 orders of magnitude
lower than the rf-modified spin-relaxation rate at the trap center.
Consequently Fig.\ \ref{fig:87Rb2-Brf0.5}(b) is totally dominated by spin
relaxation. In this case, however, the spin relaxation itself shows more
complicated structure as a function of $B$; the dashed lines in Fig.\
\ref{fig:39K_5e-1G-rf-dressed-loss_lmax0vs2}(b) show the rf-free spin
relaxation rates for $(1,1)$+$(1,1)$, $(1,1)$+$(1,0)$, $(1,1)$+$(1,-1)$ and
$(1,0)$+$(1,-1)$ collisions. As for $^{39}$K, the losses for rf-dressed states
approach those for rf-free $(1,1)$+$(1,1)$ at high magnetic field, but around
the trap centre there are also contributions from other components of the
wavefunction of the rf-dressed atomic states.

The rf-induced loss rate depends strongly on the singlet and triplet scattering
lengths $a_{\rm s}$ and $a_{\rm t}$. In order to explore this, we have carried
out $L_{\rm max}=0$ calculations on a set of potentials modified at short range
to allow adjustment of $a_{\rm s}$ and $a_{\rm t}$. We retained the functional
forms of the potential curves of Strauss {\em et al.}\ \cite{Strauss:2010}, but
modified the short-range matching point $R_{\rm SR}$ to 3.5~\AA\ for the
singlet potential and to 5.6~\AA\ for the triplet potential in order to provide
sufficient flexibility to adjust the scattering lengths through a complete
cycle. We then adjusted the short-range power $N_{\rm SR}$ to obtain modified
potentials with different scattering lengths, maintaining continuity of the
potential and its derivative at $R_{\rm SR}$ as described in ref.\
\cite{Groebner:KCs:2017}.

Contour plots of the resulting rf-induced peak loss rates $k_2^{\rm max}$ and
the corresponding real part of the scattering length $a_{\rm rf}$ (for
collisions of rf-dressed atoms) are shown in Figure \ref{fig:contour},
calculated at the trap center for $B_{\rm rf}=0.5$~G. Since the possible
singlet and triplet scattering lengths range from $-\infty$ to $+\infty$, the
loss rate is displayed as a function of two phases, defined as the fractional
parts of the quantum numbers at dissociation $v_{\rm D,s}$ and $v_{\rm D,t}$
for the singlet and triplet states. These each map onto the corresponding
scattering length according to
\begin{equation}
a = \bar{a}\left[1-
\tan\left(v_{\rm D}+\textstyle\frac{1}{2}\right)\pi\right],
\label{eq:avD}
\end{equation}
where $\bar{a} = 0.477 988\ldots (2\mu C_6/\hbar^2)^{-1/4}$ is the mean
scattering length of Gribakin and Flambaum \cite{Gribakin:1993} and $C_6$ is
the leading long-range dispersion coefficient. For $^{87}$Rb, $\bar{a}=78.95$
bohr. The mapping between scattering length and $v_{\rm D}$ is shown for
$^{87}$Rb in the top panel of Fig.\ \ref{fig:contour}.

\begin{figure}[t]
\includegraphics[width=1\columnwidth]{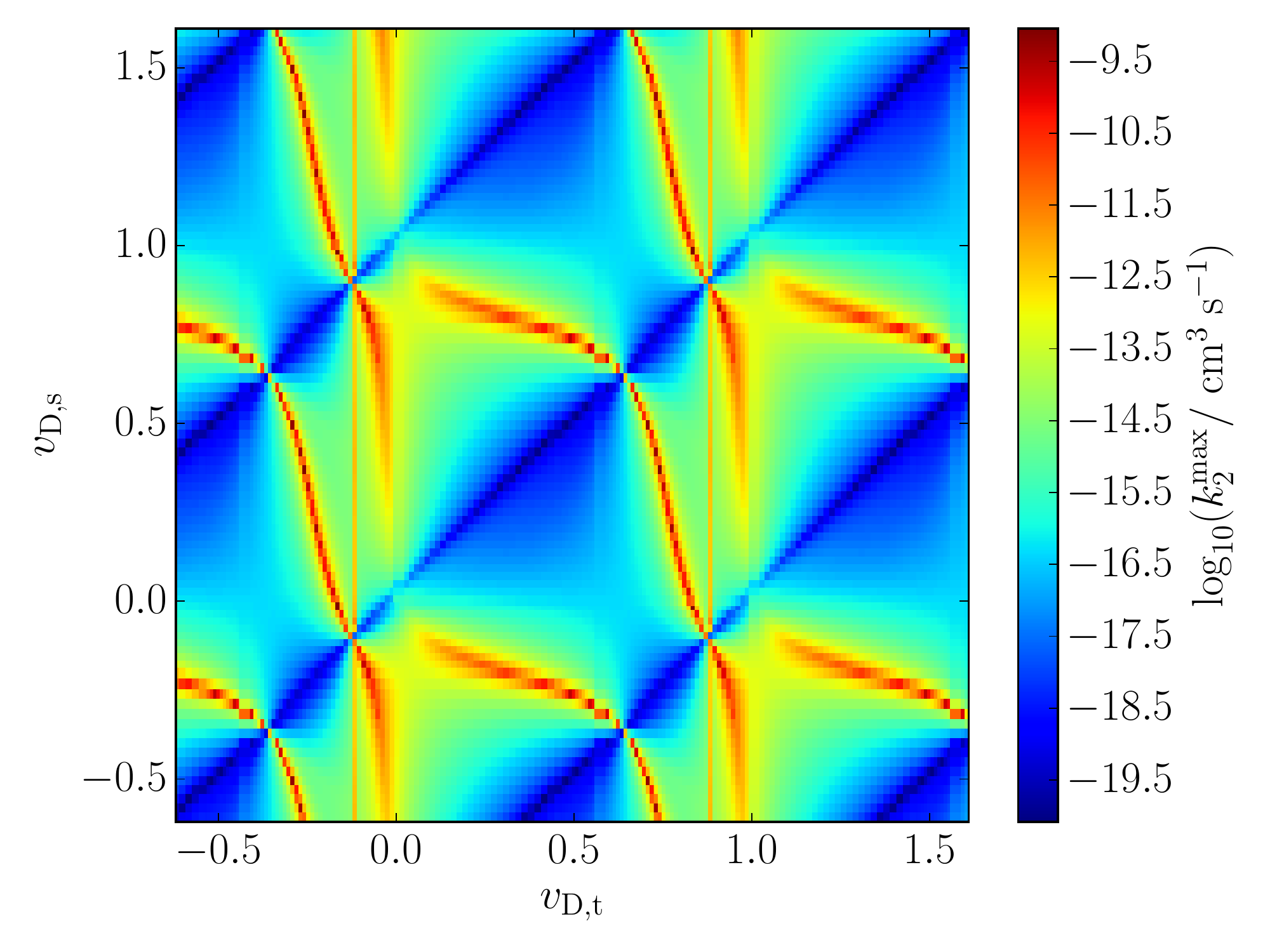}
\caption{Contour plot of the rate coefficient for rf-induced loss at the trap
center, for adiabatically trapped $f=1$ states of an artificial atom with a
hyperfine splitting 0.7 times that of $^{87}$Rb. All other quantities are the
same as for Fig.\ \ref{fig:contour}.} \label{fig:contour-art}
\end{figure}

Figure \ref{fig:contour} shows that $k_2^{\rm max}$ varies by more than 10
orders of magnitude as a function of the singlet and triplet scattering
lengths. Both $k_2^{\rm max}$ and $a_{\rm rf}$ depend only on the fractional
parts of $v_{\rm D}$ for the singlet and triplet states (and hence on $a_{\rm
s}$ and $a_{\rm t}$), as indicated by the repeating patterns in Fig.\
\ref{fig:contour}. The most striking feature of Fig.\ \ref{fig:contour}(a) is a
deep diagonal trough in the rf-induced loss rate when $v_{\rm D,s}\approx
v_{\rm D,t}$ ($a_{\rm s}\approx a_{\rm t}$), with no corresponding feature in
$a_{\rm rf}$. Superimposed on this are peaks in $k_2^{\rm max}$ and poles in
the corresponding $a_{\rm rf}$. These are of three different types. First,
there are near-vertical bands near integer values of $v_{\rm D,t}$,
corresponding to $|a_{\rm t}|=\infty$. These are entrance-channel resonances;
they occur near integer values of $v_{\rm D,t}$ because the incoming channel is
mostly triplet in character. Secondly, there is a Feshbach resonance due to a
closed channel with excited hyperfine character ($f=2$ here), which produces
curving bands of peaks in $k_2^{\rm max}$ that cross the vertical bands near
$v_{\rm D,s}=0.1$. Lastly, there is an additional Feshbach resonance that
produces very narrow vertical bands of peaks near $v_{\rm D,t}=0.3$; these
probably arise from pure triplet states that exist at the $(f_1,f_2)=(1,2)$
threshold in the absence of rf and magnetic fields.

To explore the dependence of the pattern on hyperfine splitting, we have
repeated the calculations on a series of artificial systems with the $^{87}$Rb
hyperfine splitting reduced from its real value, using the same set of
interaction potentials. The results with the hyperfine splitting at 70\% of its
real value are shown in Fig.\ \ref{fig:contour-art}. The general form of the
contour plot is unchanged, with a deep trough around $v_{\rm D,s}\approx v_{\rm
D,t}$ ($a_{\rm s}\approx a_{\rm t}$) and peaks around vertical bands at integer
values of $v_{\rm D,t}$. As expected, however, the Feshbach peaks have shifted.
They now display distinct {\em avoided} crossings with the vertical bands of
peaks. For some values of the hyperfine splitting, the crossings are so
strongly avoided that the vertical bands near integer $v_{\rm D,t}$ are barely
identifiable.

\begin{figure*}[t]
\includegraphics[width=2\columnwidth]{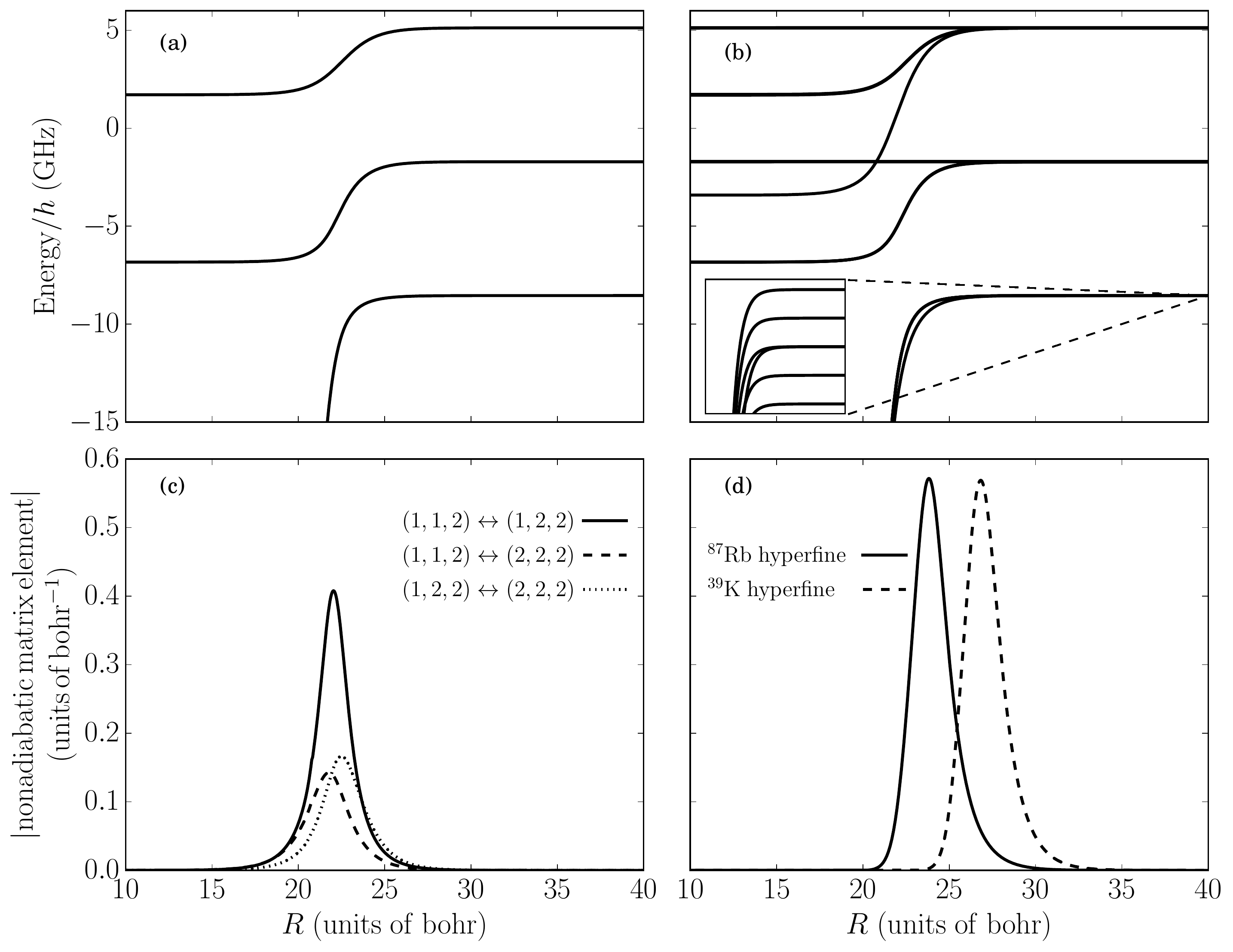}
\caption{Adiabats (eigenvalues of the Hamiltonian of Eq.\  \ref{eq:ham-pair} at
fixed $R$) with respect to a pure triplet curve (a) for field-free collisions
with $(f_1,f_2,F=2)$; (b) for collisions of rf-dressed atoms with $f=1$, at the
trap center with rf field frequency 3 MHz and strength $B_{\rm rf}=0.5$\,G, for
$M_{\rm tot}=0$. (c) nonadiabatic matrix elements between (1,1,2), (1,2,2) and
(2,2,2) in (a); (d) nonadiabatic matrix element between the uppermost of the six
$(f_1,f_2,M_{\rm tot})=(1,1,0)$ rf-dressed states and the next-highest state.}
\label{fig:nonad1}
\end{figure*}

The trough observed here for rf-induced inelastic collisions when $a_{\rm
s}\approx a_{\rm t}$ has a similar origin to that found for spin-exchange
collisions. Myatt {\em et al.}\ \cite{Myatt:1997} measured a very low rf-free
spin-exchange rate in dual Bose-Einstein condensates of $^{87}$Rb in
$(f,m_f)=(1,-1)$ and (2,2) states. Julienne {\em et al.}\ \cite{Julienne:1997}
explained this using an adiabatic model of the collision dynamics
\cite{Mies:1996} where the Hamiltonian of Eq.\ \ref{eq:ham-pair} (without rf)
is diagonalized at each value of the interatomic distance $R$. For $^{87}$Rb,
the exchange splitting between the singlet and triplet curves is comparable to
the hyperfine splitting around $R_{\rm X} = 22$ bohr. Inside this distance the
adiabatic states are essentially pure singlet and triplet states, whereas
outside it they are described by atomic quantum numbers $(f,m_f)$. Julienne
{\em et al.}\ \cite{Julienne:1997} considered scattering at zero magnetic
field, with $f_1$ and $f_2$ coupled to give a resultant $F$. The resulting
rf-free adiabats for $F=2$ are shown in Fig.\ \ref{fig:nonad1}(a), with respect
to the pure triplet interaction potential. The nonadiabatic matrix elements
$\langle i | d/dR | j\rangle$ between the states that are asymptotically
$(f_1,f_2,F)=(1,1,2), (1,2,2)$ and (2,2,2) are shown in Fig.\
\ref{fig:nonad1}(c). The overall magnitude is usefully characterized by the
integral
\begin{equation}
D_{ij} = \int \left\langle i \left| \frac{d}{dR} \right| j\right\rangle\,dR,
\end{equation}
which is $\pi/2$ for a complete avoided crossing, and 1.05, 0.55 and 0.49 for
the three couplings in Fig.\ \ref{fig:nonad1}(c). The adiabats and nonadiabatic
couplings are independent of the singlet and triplet scattering lengths.
However, Julienne {\em et al.} \cite{Julienne:1997} argued that, when $a_{\rm
s}\approx a_{\rm t}$, the radial wavefunctions $R^{-1}\chi_i(R)$ and
$R^{-1}\chi_j(R)$ in the incoming (1,2,2) and inelastic outgoing (1,1,2)
channels are in phase around $R_{\rm X}$. This minimizes the matrix element
that controls inelastic scattering,
\begin{equation}
-\frac{\hbar^2}{2\mu} \int \chi_i(R)^*
\left\langle i \left| \frac{d}{dR} \right| j\right\rangle \frac{d}{dR}\,\chi_j(R)\,dR.
\end{equation}
Figures \ref{fig:nonad1}(a) and (c) may be compared with the adiabats for
collisions of rf-dressed $^{87}$Rb atoms in $f=1$ states at the trap center
(4.27~G), which are shown in Fig.\ \ref{fig:nonad1}(b), and the nonadiabatic
matrix element from the uppermost of the six $(f_1,f_2,M_{\rm tot})=(1,1,0)$
states to the next-highest state, which are shown in Fig.\ \ref{fig:nonad1}(d).
The nonadiabatic matrix element again peaks around 22 bohr for $^{87}$Rb; the
mechanism is similar to that for spin exchange, and the overall inelastic
coupling is smallest when $a_{\rm s}\approx a_{\rm t}$, producing the diagonal
troughs seen in Figs.\ \ref{fig:contour} and \ref{fig:contour-art}.

The actual singlet and triplet scattering lengths for $^{87}$Rb are indicated
by a cross on Fig.\ \ref{fig:contour}(a). This shows that $^{87}$Rb is special
in two different ways. Not only are its singlet and triplet scattering lengths
quite similar, but their actual values correspond to $v_{\rm D}\approx 0.5$ and
lie well away from the peaks due to Feshbach resonances. The value of $k_2^{\rm
max}$ at the deepest point in the trough in Fig.\ \ref{fig:contour}(a) is about
$k_2^{\rm max}=3.6\times 10^{-20}$ cm$^3$\,s$^{-1}$, which is not far from the
value of $1.25 \times 10^{-20}$ cm$^3$\,s$^{-1}$ obtained for $^{87}$Rb on the
potentials of ref.\ \cite{Strauss:2010}.

\begin{figure}[t]
\includegraphics[width=1\columnwidth]{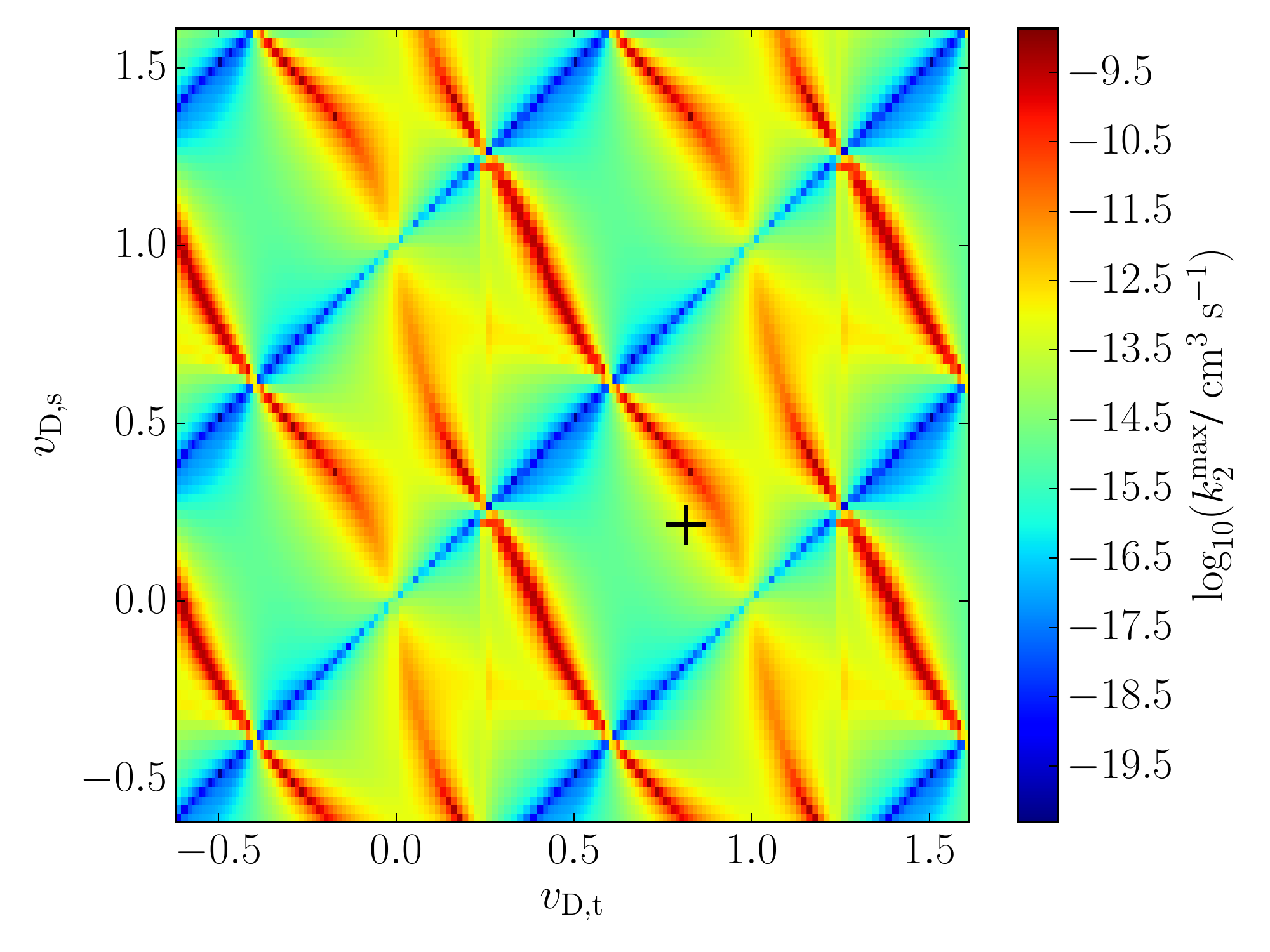}
\caption{Contour plot of the rate coefficient for rf-induced loss at the trap
center, for adiabatically trapped $f=1$ states of an artificial atom with the
mass of $^{87}$Rb with the hyperfine splitting of $^{39}$K. All other
quantities are the same as for Fig.\ \ref{fig:contour}.} \label{fig:contour-k}
\end{figure}

Figure \ref{fig:contour-k} shows a contour plot similar to Fig.\
\ref{fig:contour-art} but with the hyperfine splitting of $^{39}$K (462 MHz).
The structure is similar, with a Feshbach resonance avoided-crossing with
vertical bands of peaks at integer $v_{\rm D,t}$, though the resonances are
distinctly wider than in Figs.\ \ref{fig:contour}(a) and \ref{fig:contour-art}.
The actual scattering lengths of $^{39}$K are shown as a black cross; the value
of $k_2^{\rm max}$ at this point is $5.3\times 10^{-14}$ cm$^3$\,s$^{-1}$,
which may be compared with $k_2^{\rm max}=6.33 \times 10^{-14}$
cm$^3$\,s$^{-1}$ from the calculation with the mass and interaction potentials
for $^{39}$K in section \ref{sec:k2}.

\begin{figure}[t]
\includegraphics[width=\columnwidth]{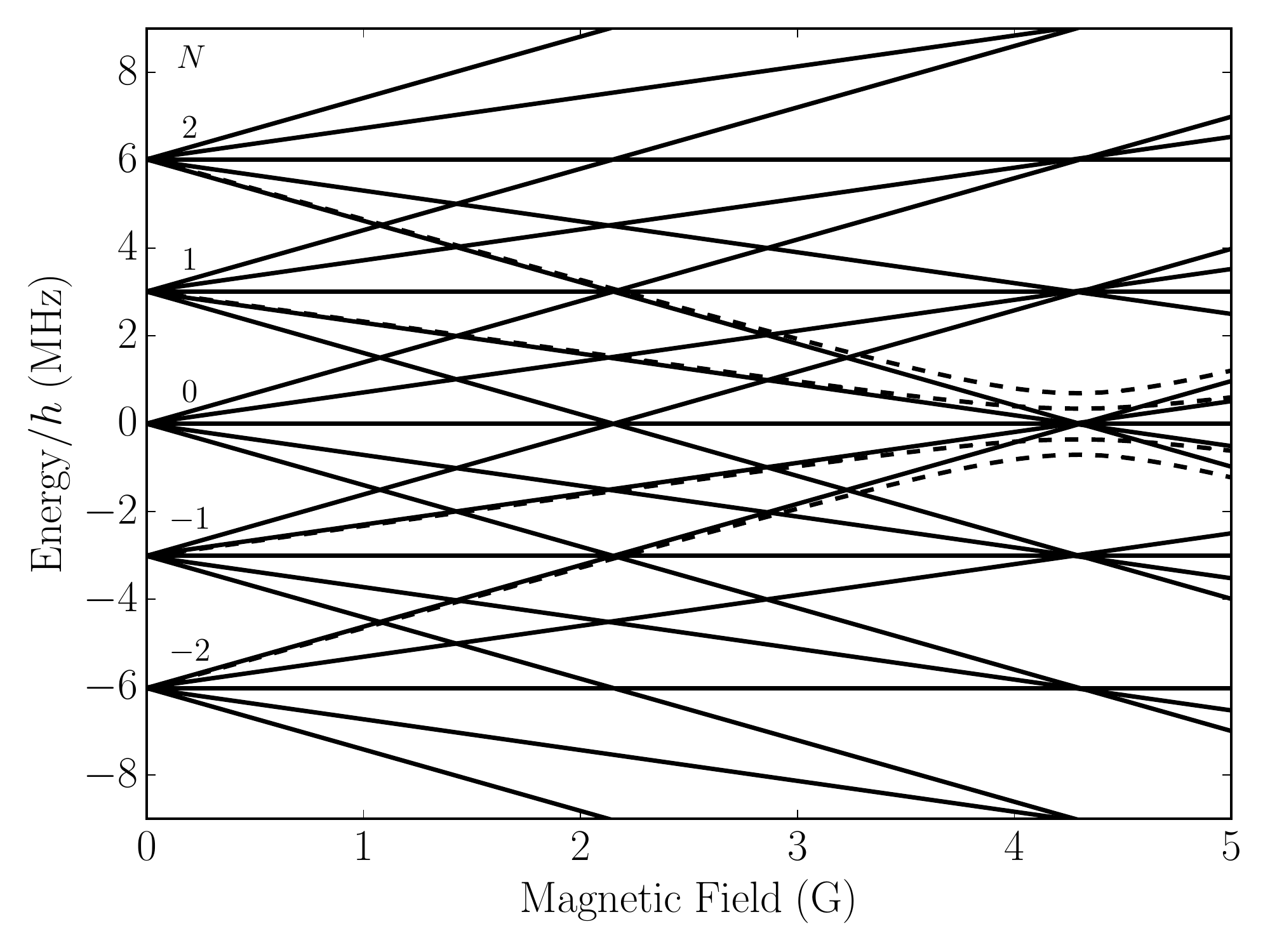}
\caption{rf-dressed atomic levels of $f=2$ states of $^{87}$Rb for frequency
$3.0$ MHz and photon numbers $N=-2$, -1, 0, 1 and 2, shown with respect to the
energy of the $f=2$, $m_f=0$ state for $N=0$. Solid lines show levels for zero
rf intensity and dashed lines show levels for $B_{\rm rf}=0.5$\,G with $M_{\rm
tot}=0$. Atoms can be trapped at the minimum in the uppermost dashed curve.}
\label{fig:Rbatomicf2}
\end{figure}

\begin{figure}[t]
\includegraphics[width=\columnwidth]{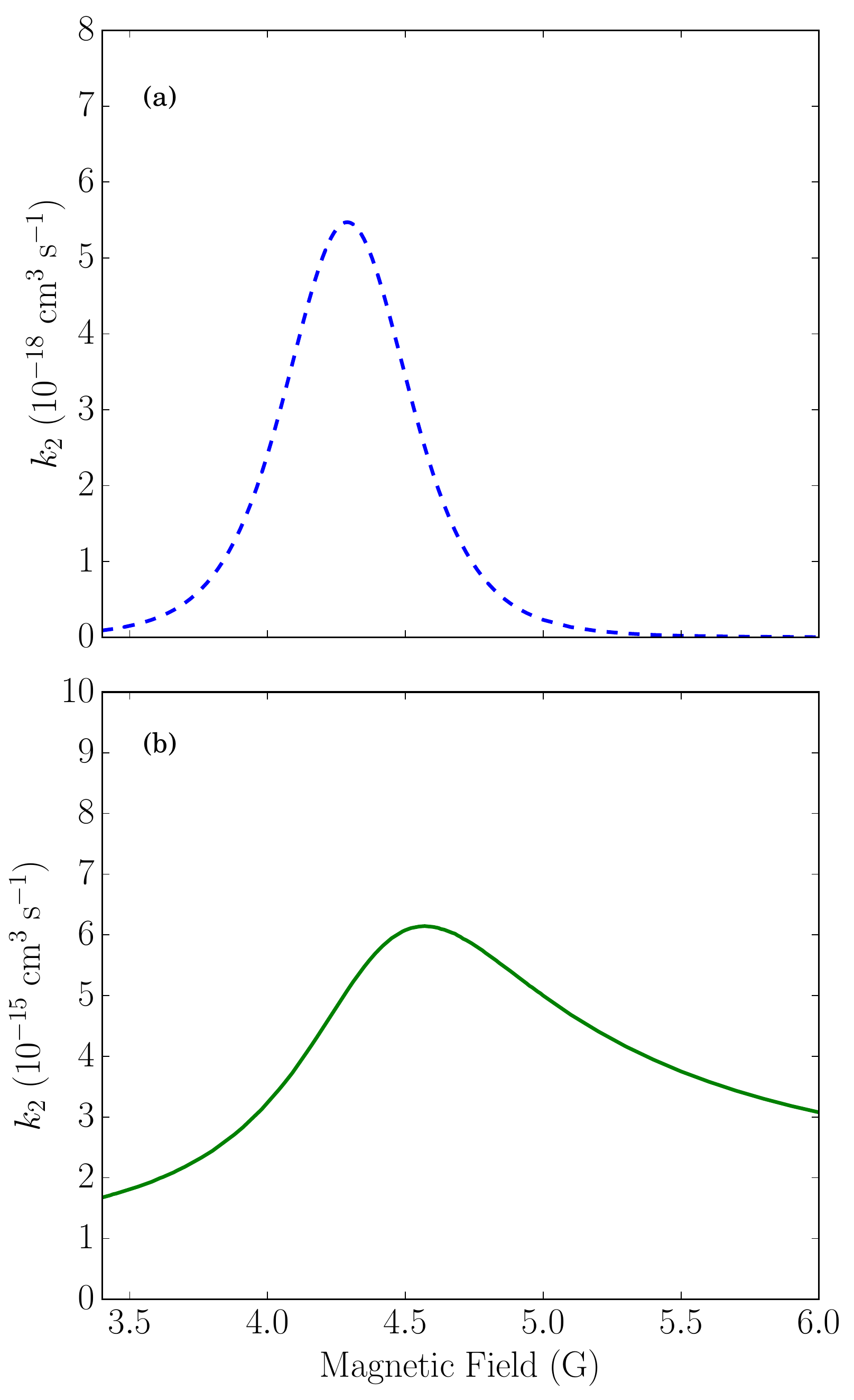}
\caption{Rate coefficient for inelastic loss of adiabatically trapped $^{87}$Rb
in $f=2$ as a function of magnetic field with $\nu = 3.0$ MHz and $B_{\rm rf} =
0.5$ G. (a) Calculation of rf-induced loss, using $L_{\rm max}=0$. (b)
Calculation including spin relaxation, using $L_{\rm max}=2$.}
\label{fig:87Rb2-f2-Brf0.5}
\end{figure}

\subsection{Inelastic collisions of rf-dressed $f=2$ states}

A somewhat different case occurs for atoms in $f=2$ states. Here there are 5
photon-dressed atomic states that cross as a function of magnetic field, as
shown for $^{87}$Rb in Fig.\ \ref{fig:Rbatomicf2}. It requires a minimum of 5
rf-free states (with photon numbers $N$ from $-2$ to 2) to describe a single
trapped atom, and describing two such atoms requires photon numbers from $-4$
to 4. The coupled-channel calculation is thus computationally considerably more
expensive. Nevertheless, the principles are exactly the same and rate
coefficients for inelastic loss can again be obtained from $\beta$, the
imaginary part of the complex scattering length, for atoms initially at the
highest rf-dressed threshold.

\begin{figure}[t]
\includegraphics[width=1\columnwidth]{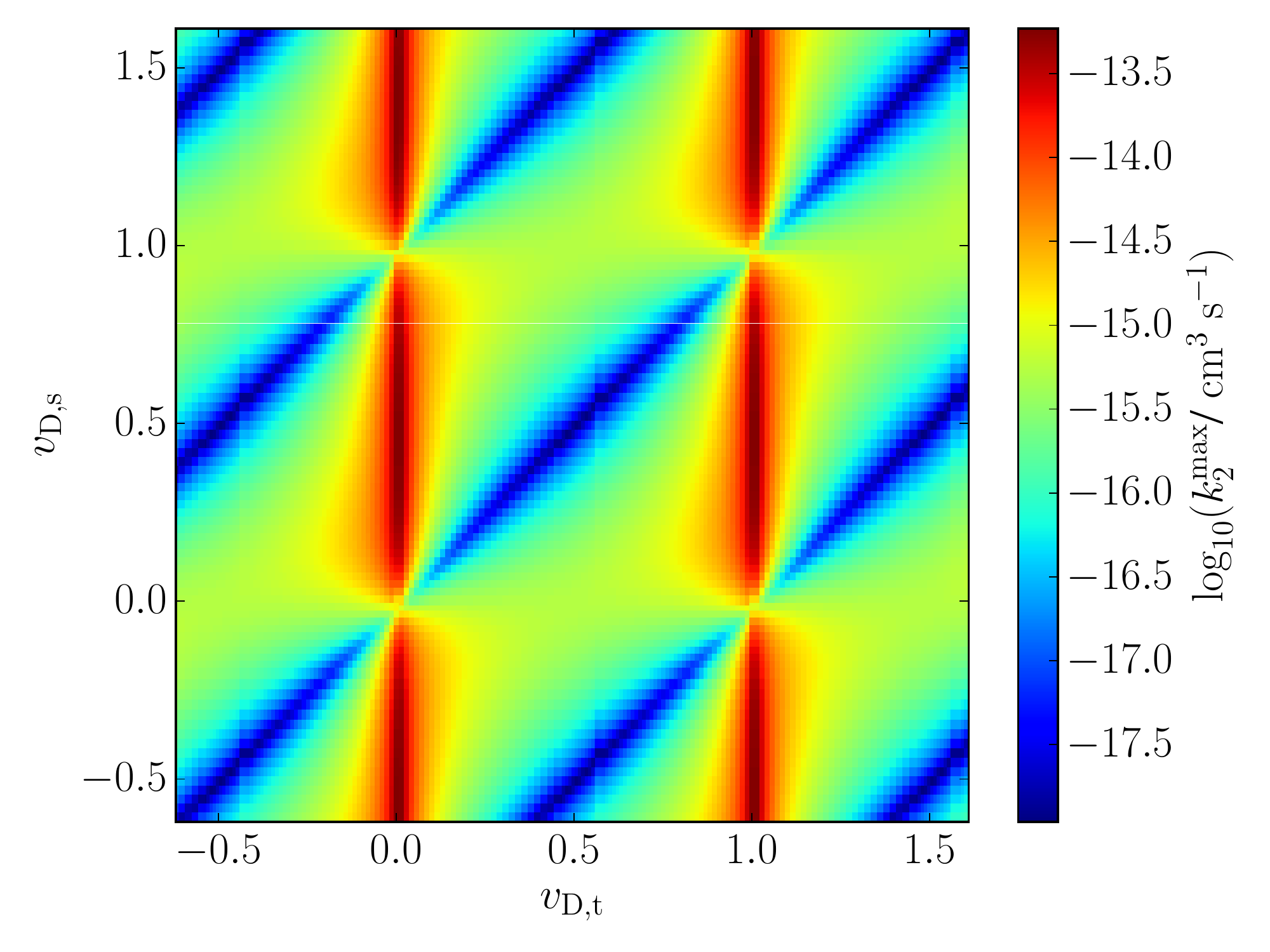}
\caption{Contour plot of the rate coefficient for rf-induced loss at the trap
center, for adiabatically trapped $f=2$ states of $^{87}$Rb. All other
quantities are the same as for Fig.\ \ref{fig:contour}. }
\label{fig:contour-f2}
\end{figure}

Figure \ref{fig:87Rb2-f2-Brf0.5} shows the rate coefficient for inelastic loss
for $^{87}$Rb in $f=2$, as a function of magnetic field near the trap center.
As before, Fig.\ \ref{fig:87Rb2-f2-Brf0.5}(a) shows the rf-induced loss, from a
calculation with $L_{\rm max}=0$, while Fig.\ \ref{fig:87Rb2-f2-Brf0.5}(b)
shows the loss including spin relaxation, from a calculation with $L_{\rm
max}=2$. The rf-induced loss rate is about a factor of 400 larger than for
$^{87}$Rb in $f=1$, but it is still much lower than the loss rate due to spin
relaxation. Once again this illustrates the special properties of $^{87}$Rb.

\begin{figure}[t]
\includegraphics[width=1\columnwidth]{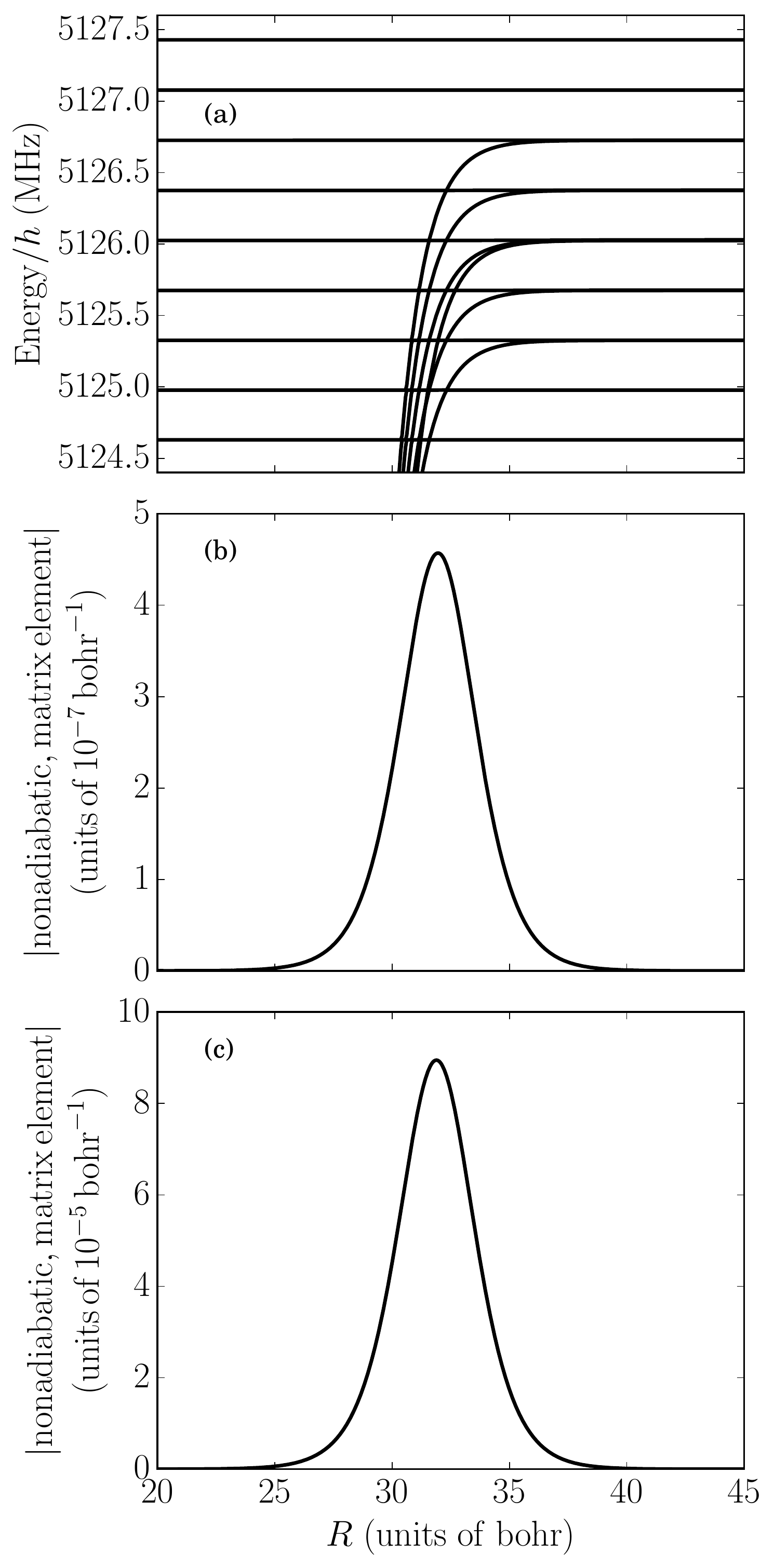}
\caption{(a) Adiabats (eigenvalues of the Hamiltonian of Eq.\ \ref{eq:ham-pair}
at fixed $R$) for collisions of field-dressed $^{87}$Rb atoms in $f=2$ states,
with respect to a pure triplet curve, for $M_{\rm tot}=0$. (b) nonadiabatic
matrix elements between the top two adiabatic states for $^{87}$Rb. (c)
nonadiabatic matrix elements between the top two adiabatic states with the
hyperfine splitting reduced to the value for $^{39}$K.} \label{fig:adiabats-f2}
\end{figure}

It is notable that the rf-induced loss rate for $f=2$ is far lower than the
rf-free spin-exchange rates for (2,0)+(2,0) and (2,1)+(2,$-1$) collisions,
which are $1.73 \times 10^{-13}$ and $1.25 \times 10^{-13}$ cm$^3$\,s$^{-1}$
respectively. This is true even though the wavefunction for the rf-dressed
atomic state includes substantial amounts of (2,0) and (2,1) near the trap
center. It may be again rationalized by considering adiabatic curves obtained
by diagonalizing the Hamiltonian of Eq.\ \ref{eq:ham-pair} at each value of the
interatomic distance $R$. For rf-free collisions of two $f=2$ atoms, there are
contributions from $F=0$, 2 and 4. The inelasticity is dominated by $F=2$, for
which the adiabats and nonadiabatic matrix elements were shown in Fig.\
\ref{fig:nonad1}(a) and (c). Figure \ref{fig:adiabats-f2}(a) shows the adiabats
for collisions of rf-dressed $^{87}$Rb atoms in $f=2$ states at the trap
center, and Fig.\ \ref{fig:adiabats-f2}(b) shows the corresponding nonadiabatic
matrix elements between the uppermost and next-highest state. The nonadiabatic
coupling is quite different from the previous cases: there is no feature around
22 bohr, and instead the matrix element peaks around 32 bohr, where the
difference between the singlet and triplet curves is comparable with the
splittings $\Delta_{\rm rf}$ due to rf dressing. The coupling is far weaker
than in the cases shown in Fig.\ \ref{fig:nonad1}. The integral $D_{ij}$ over
the nonadiabatic coupling is only $1.96\times10^{-6}$, as compared to $\pi/2$
for a completed avoided crossing.

We have also investigated the dependence of $k_2^{\rm max}$ for the $f=2$
states of $^{87}$Rb$_2$ on the singlet and triplet scattering lengths. The
resulting contour plot is shown in Fig.\ \ref{fig:contour-f2}. It has a
considerably simpler structure than Figs.\ \ref{fig:contour},
\ref{fig:contour-art} and \ref{fig:contour-k}, because the atoms are both in
their upper hyperfine state and there are no closed channels that can cause
Feshbach resonances. The only features are a diagonal trough for $v_{\rm
D,s}\approx v_{\rm D,t}$ ($a_{\rm s}\approx a_{\rm t}$) and a near-vertical
band of maxima where $v_{\rm D,t}$ is close to an integer. These have the same
causes as discussed for $f=1$ above.

\begin{figure}[t]
\includegraphics[width=1\columnwidth]{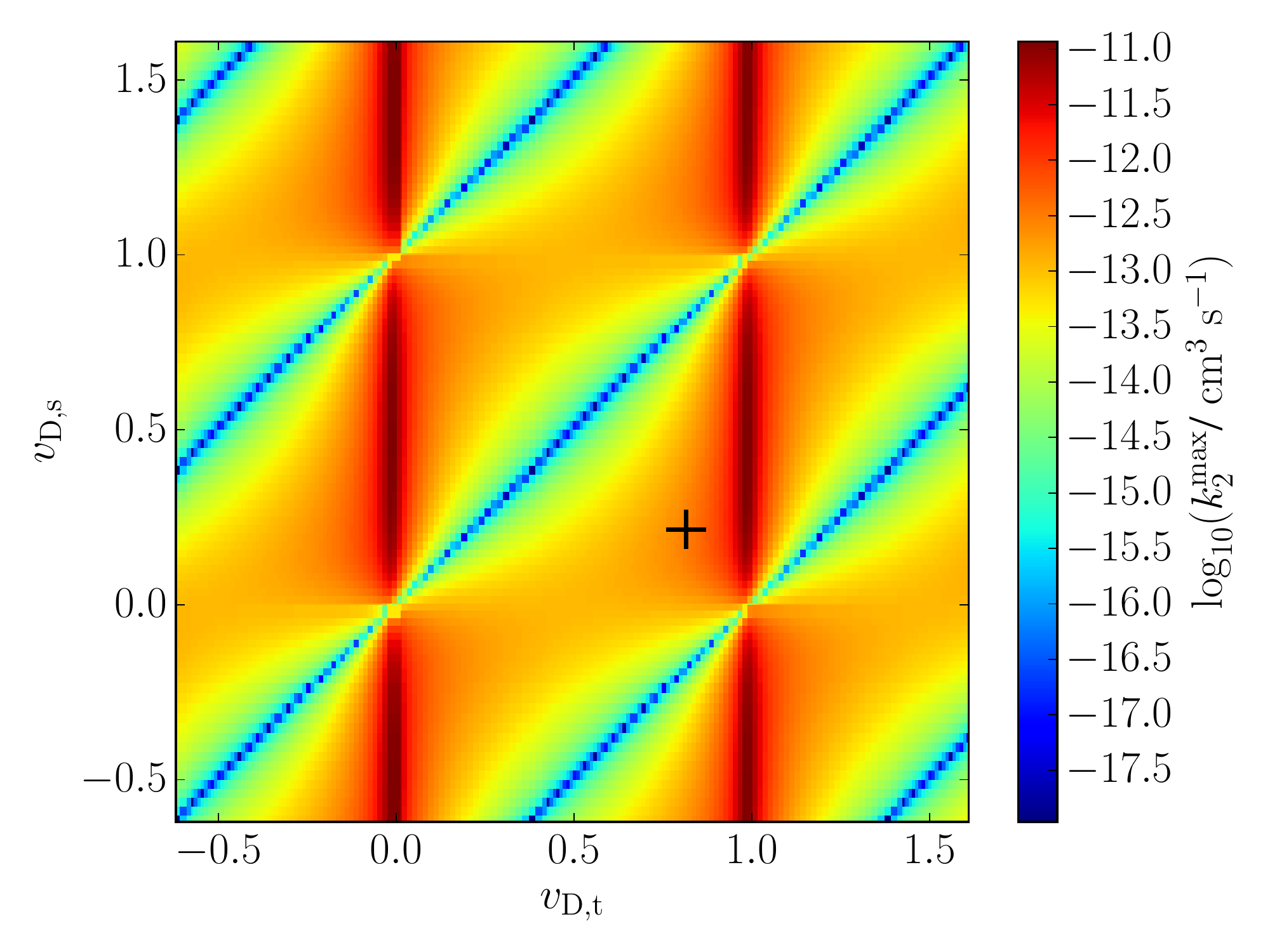}
\caption{Contour plot of the rate coefficient for rf-induced loss at the trap
center, for adiabatically trapped $f=2$ states of an artificial atom with the
mass of $^{87}$Rb but the hyperfine splitting of $^{39}$K. All other quantities
are the same as for Fig.\ \ref{fig:contour}. } \label{fig:contour-f2-K}
\end{figure}

\begin{figure}[t]
\includegraphics[width=\columnwidth]{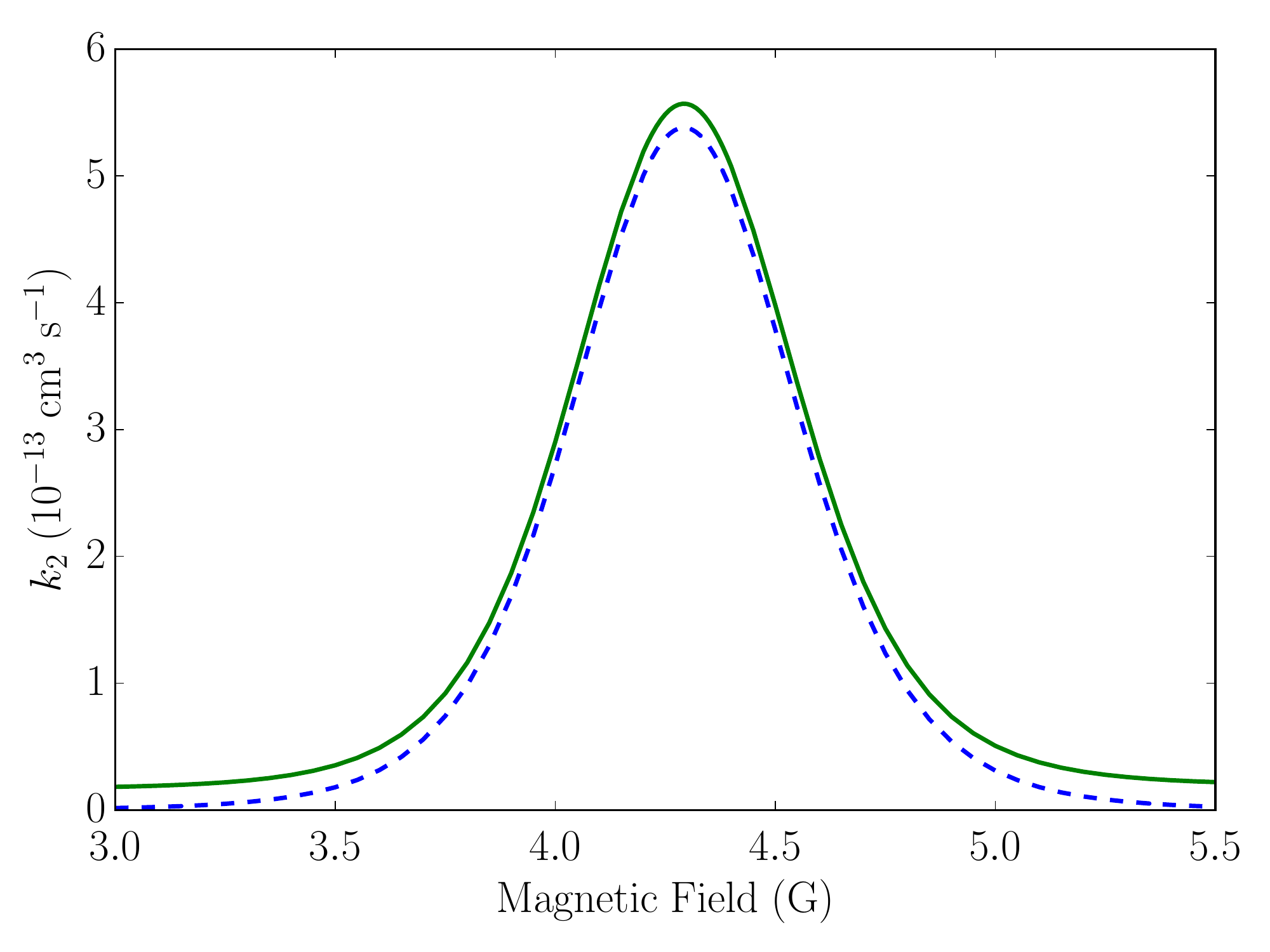}
\caption{Rate coefficient for inelastic loss of adiabatically trapped $^{39}$K
in $f=2$ as a function of magnetic field with $\nu = 3.0$ MHz and $B_{\rm rf} =
0.5$ G. Results are shown including spin relaxation, using $L_{\rm max}=2$
(solid, green) and for rf-induced loss alone, using $L_{\rm max}=0$ (dashed,
blue).} \label{fig:39K2-f2-Brf0.5}
\end{figure}

For atoms trapped in their upper hyperfine state, with no Feshbach resonances,
the dependence of $k_2^{\rm max}$ on $a_{\rm s}$ and $a_{\rm t}$ may be
expected to resemble Fig.\ \ref{fig:contour-f2} qualitatively for all atoms.
However, there is a strong overall dependence on the hyperfine coupling
constant. To illustrate this, we have repeated the calculations shown in Fig.\
\ref{fig:contour-f2} with the hyperfine coupling constant set to the value for
$^{39}$K but the reduced mass retained at the value for $^{87}$Rb. The results
are shown in Fig.\ \ref{fig:contour-f2-K}. It may be seen that the general
structure of peaks and troughs is unchanged, but the peaks are about a factor
of 200 higher for the smaller hyperfine splitting of $^{39}$K (462 MHz) than
for that of $^{87}$Rb (6,834 MHz). This effect may also be traced to the
effects of nonadiabatic transitions. The adiabats for $^{39}$K are similar to
those shown for $^{87}$Rb in Fig.\ \ref{fig:adiabats-f2}(a). However, the
nonadiabatic matrix element, shown in Fig.\ \ref{fig:adiabats-f2}(c), is a
factor of 200 larger than for $^{87}$Rb, with integral
$D_{ij}=3.8\times 10^{-4}$. The nonadiabatic matrix element
reflects the amount of singlet character in the wavefunction of the rf-dressed
atomic pair at long range; this in turn depends on the degree of mixing of
$f=1$ and $f=2$ states in the magnetic field, which increases as the hyperfine
splitting decreases.

The specific case of $^{39}$K in rf-dressed $f=2$ states is of interest. Figure
\ref{fig:39K2-f2-Brf0.5} shows $k_2$ as a function of magnetic field from
calculations with $L_{\rm max}=0$ and 2, using the potentials of ref.\
\cite{Falke:2008}. It may be seen that, as for $^{39}$K in $f=1$, the
rf-induced loss dominates the loss due to rf-modified spin relaxation. The rate
coefficient peaks at $k_2^{\rm max}=5.38\times 10^{-13}$ cm$^3$\,s$^{-1}$. The
rf-induced loss is about 5 orders of magnitude faster than for $^{87}$Rb, and
again more typical. The value is comparable to the one from Fig.\
\ref{fig:contour-f2-K} at the values of $v_{\rm D,s}$ and $v_{\rm D,t}$ for
$^{39}$K, shown with a black cross, which is $k_2^{\rm max}=2.34\times
10^{-13}$ cm$^3$\,s$^{-1}$. The difference between these two values arises
because the calculations in Fig.\ \ref{fig:contour-f2-K} used the reduced mass
and interaction potentials for $^{87}$Rb rather than $^{39}$K.

An atom for which $f=2$ is not the highest state, such as $^{85}$Rb, may be
expected to show more complex behavior than Fig.\ \ref{fig:contour-f2}, with
Feshbach resonances reappearing at some values of $v_{\rm D,s}$ and $v_{\rm
D,t}$ ($a_{\rm s}$ and $a_{\rm t}$). We will investigate this in future work.

\section{Conclusions}

Cold atoms in radiofrequency-dressed traps may undergo inelastic collisions by
mechanisms that do not occur in the absence of an rf field. These collisions
may lead to trap loss. We have investigated inelastic losses for alkali-metal
atoms in rf-dressed traps, using coupled-channel scattering calculations on
accurate potential energy surfaces. We have explored the dependence of the loss
rates on singlet and triplet scattering lengths, hyperfine splittings and the
strength of the rf field.

There are two components of the inelastic loss. One is due to spin-relaxation
collisions, driven by the dipolar interaction between the electron spins of the
two atoms. This component exists even in the absence of rf dressing, but is
generally fairly small, both because the dipolar interaction is weak and
because there is a centrifugal barrier in the outgoing channel. It is modified
near the trap center because the rf-dressed states are mixtures of different
spin states, and drops to zero on the low-field side of the trap, where the
adiabatically trapped state correlates with lowest state in the magnetic field.
The second component, which we refer to as rf-induced loss, is potentially
stronger; the inelastic collisions are driven by the difference between the
singlet and triplet interaction potentials, and there is no centrifugal barrier
in the outgoing channel.

For $^{87}$Rb in $f=1$ states, the calculated rate coefficient for rf-induced
loss is very small. We obtain $k_2^{\rm max}=1.25 \times 10^{-20}$
cm$^3$\,s$^{-1}$ at the trap center for an rf field strength $B_{\rm
rf}=0.5$~G. This is much smaller than the rf-modified spin-relaxation loss rate
coefficient for $^{87}$Rb.

We have explored the dependence of the rf-induced loss rate on the singlet and
triplet scattering lengths $a_{\rm s}$ and $a_{\rm t}$, and find that it can
change by 10 orders of magnitude as the scattering lengths are varied. It is
generally small when $a_{\rm s} \approx a_{\rm t}$, but may be enhanced by
resonances of two different types. $^{87}$Rb is a very special case: not only
is $a_{\rm s}$ very similar to $a_{\rm t}$, but their actual values are such
that there is no enhancement by either type of resonance. Other alkali-metal
atoms will generally have much larger rf-induced loss rates. For $^{39}$K,
which is a more typical case, we obtain $k_2^{\rm max}=6.33 \times
10^{-14}$ cm$^3$\,s$^{-1}$ for $B_{\rm rf}=0.5$~G. This is much larger than
the rf-modified spin-relaxation loss rate, and 6 orders of magnitude larger
than for $^{87}$Rb. The rf-induced loss rate at the trap center increases at
lower rf field strengths.

We have also investigated rf-induced loss for alkali-metal atoms in their upper
hyperfine states, $f=2$ for $^{87}$Rb and $^{39}$K. These losses are also small
when $a_{\rm s} \approx a_{\rm t}$. In this case there are no Feshbach
resonances, but the loss rates may still be enhanced by entrance-channel
effects when $|a_{\rm t}|$ is large. The rf-induced loss rates also depend
strongly on the atomic hyperfine splitting, increasing as the hyperfine
splitting decreases because of mixing of atomic $f$ states by the magnetic
field.

\acknowledgments

We are grateful to Prof. C. J. Foot and Mr.\ E. Bentine for interesting us in
this project and to them, Dr.\ M. D. Frye and Dr.\ C. R. Le Sueur for valuable
discussions. This work has been supported by the UK Engineering and Physical
Sciences Research Council (Grants No. ER/I012044/1, EP/N007085/1 and
EP/P01058X/1 and a Doctoral Training Partnership with Durham University).

\bibliography{../all}
\end{document}